\documentclass[9pt,twocolumn,twoside]{osajnl}

\journal{ol}
\setboolean{shortarticle}{true} 

\title{Quadratic Soliton Combs in Doubly-Resonant Second-Harmonic Generation}

\author[1,2,*]{Tobias Hansson}
\author[3,4]{Pedro Parra-Rivas}
\author[2]{Martino Bernard}
\author[3]{François Leo}
\author[4]{Lendert Gelens}
\author[2,5]{Stefan Wabnitz}

\affil[1]{Department of Physics, Chemistry and Biology, Linköping University, SE-581 83 Linköping, Sweden}
\affil[2]{Dipartimento di Ingegneria dell'Informazione, Universit\`a di Brescia, via Branze 38, 25123 Brescia, Italy}
\affil[3]{OPERA-photonics, Université libre de Bruxelles, 50 Avenue F. D. Roosevelt, CP 194/5, B-1050 Bruxelles, Belgium}
\affil[4]{Laboratory of Dynamics in Biological Systems, KU Leuven Department of Cellular and Molecular Medicine, University of Leuven, B-3000 Leuven, Belgium}
\affil[5]{CNR-–INO, via Branze 38, 25123 Brescia, Italy}

\affil[*]{Corresponding author: tobias.hansson@liu.se}

\dates{Compiled \today}

\ociscodes{(190.4410) Nonlinear optics, parametric processes; (190.5530) Pulse propagation and temporal solitons; (140.4780) Optical resonators.}

\begin{abstract}
We report a theoretical investigation of quadratic frequency combs in a dispersive second-harmonic generation cavity system. We identify different dynamical regimes and demonstrate that the same system can exhibit both bright and dark localized cavity solitons in the absence of a temporal walk-off.
\end{abstract}

\setboolean{displaycopyright}{false}

\begin{document}

\maketitle

The multiwave-mixing interactions that occur when a cavity enclosed nonlinear medium is driven near resonance can result in efficient frequency conversion and the generation of broadband optical frequency combs \cite{DelHaye_2007,Pasquazi_2018}. While most investigations to date have focused on Kerr frequency combs, where the nonlinearity is due to the third-order susceptibility, it has recently been demonstrated that frequency combs can be generated also in quadratic nonlinear media \cite{Ulvila_2013,Ricciardi_2015,Mosca_2018}. Quadratic combs may operate with substantially decreased pump power compared to Kerr combs, and may also permit the direct generation of combs in spectral regions where the generation of conventional Kerr combs is difficult to achieve; e.g., because no suitable pump sources are available or because the dispersion properties of the material are not conducive to comb generation.

Here we consider the formation of quadratic combs in a dispersive second-harmonic generation (SHG) cavity system for which both the fundamental field (FF) at frequency $\omega_0$ and the second harmonic (SH) field at $2\omega_0$ are resonant \cite{Buryak_2002,Leo_2016b}. Comb generation in this system relies on the initial frequency doubling of the driven FF to create a second-harmonic wave. The SH is in turn the source of an internally pumped optical parametric oscillator (OPO) that results in the growth of subharmonic sidebands above a certain pump threshold \cite{Marte_1994,Schiller_1997}. Subsequent, cascaded three-wave mixing interactions among the different components can then result in the formation of simultaneous combs around both fundamental and SH wavelengths.

Cavity solitons are considered to be one of the most important waveforms for frequency comb applications since they correspond to broadband, coherent and mode-locked temporal pulses with a fixed repetition rate \cite{Herr_2016}. This implies that the nonlinear frequency shift is able to compensate for the dispersion so as to produce an ideal frequency comb with an equidistant comb-line spacing of a single free-spectral-range (FSR). For quadratic combs to become a viable alternative to Kerr combs, it is therefore of considerable interest to find out if, and under what circumstances, CSs and localized solutions may be generated in quadratically nonlinear systems. Building on previously developed time-domain models for SHG combs \cite{Leo_2016a,Leo_2016b,Hansson_2017}, we study the existence of mode-locked cavity solitons (CSs) and localized solutions in a dispersive SHG cavity system. Quadratic solitons are known to be present for two-dimensional (2D) diffractive SHG cavities \cite{Etrich_1997a,Etrich_1997b}, but the conditions for their existence and stability may differ for 1D dispersive systems.

\begin{figure}[htb]
\centerline{\includegraphics[width=\linewidth,trim={1.7cm .2cm .8cm .3cm},clip]{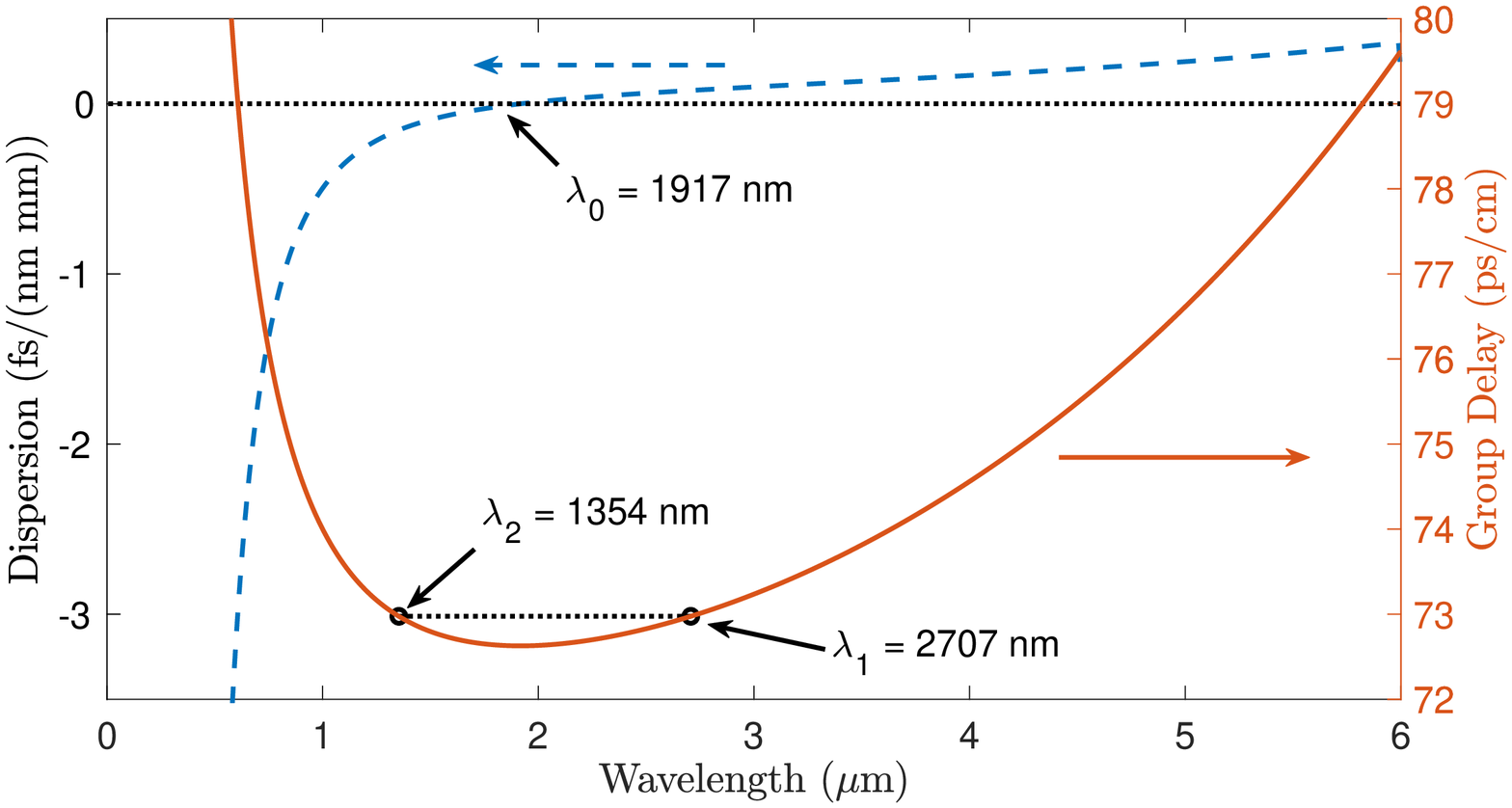}}
\caption{Wavelength dependence of dispersion and group delay for propagation along the extraordinary axis of LiNbO$_3$.}
\label{fig:LiNbO3}
\end{figure}

A distinguishing property of dispersive SHG cavities, with respect to formally equivalent diffractive systems, is that there is usually a large temporal walk-off present due to differences in group-velocity between the two field envelopes \cite{Leo_2016a}. Such a walk-off is often undesirable since it can be detrimental to the formation of localized solutions. To find a physically realizable configuration for which the walk-off vanishes, we consider the nonlinear medium to be a quasi-phase matched LiNbO$_3$ crystal \cite{Fig1} with the FF and SH wavelengths at 2707 nm and 1354 nm, respectively, on opposite sides of the zero-dispersion wavelength, as shown in Fig.~\ref{fig:LiNbO3}. We imagine that phase-matching of the crystal is achieved through periodic poling.

The dynamics of the system can, in the mean field approximation, be modeled using two coupled equations for the FF/SH fields that, in normalized form (see Ref.~\cite{Leo_2016b}), may be written as 
\begin{align}
  & \frac{\partial A}{\partial t} = \left[-(1+i\Delta_1) - i\eta_1\frac{\partial^2}{\partial\tau^2}\right]A + i\kappa B A^* + S, \label{eq:S1}\\
  & \frac{\partial B}{\partial t} = \left[- (\alpha+i\Delta_2) - d\frac{\partial}{\partial\tau} - i\eta_2\frac{\partial^2}{\partial\tau^2} \right]B+ i\kappa^*A^2,
  \label{eq:S2}
\end{align}

where $A$ and $B$ are the slowly-varying envelopes of the FF and SH field, respectively. We employ a two timescale approach, where $t$ is a slow-time variable that describes the evolution of the fields over multiple circulations, while $\tau$ is a retarded fast-time variable that describes the periodic temporal field profiles within a window with a duration of one roundtrip time. Moreover, $\alpha$ is the ratio of roundtrip losses of the SH and FF fields, while $d$ is the temporal walk-off, $\Delta_{1,2}$ are cavity detunings, $\eta_1 = sgn[k_1'']$, $\eta_2 = k_2''/|k_1''|$ are group-velocity dispersion (GVD) parameters, $\kappa$ is a nonlinear coupling constant that depends on the phase-mismatch and $S$ is the strength of the driving field. For the particular setup under consideration we use dispersion values from Fig.~\ref{fig:LiNbO3} and assume $\eta_1 = -1$, $\eta_2 = 0.5$ and $d = 0$. Here it is important that $d$ is sufficiently small, whereas there is no qualitative change for moderate variation in other parameters. For definitiveness, we assume that $\kappa = 1$, that the FF and SH losses are equal so that $\alpha = 1$, and further that the detunings are related through the condition for natural phase-matching, i.e., $\Delta_2 = 2\Delta_1$ \cite{Leo_2016b}. The dynamics is then dependent only on the specific pump settings that are experimentally accessible through changes to the driving amplitude $S$ and the frequency detuning $\Delta_1$. 

Equations (\ref{eq:S1}-\ref{eq:S2}) have a set of homogeneous mixed-mode steady-state solutions $A_0 = S(\alpha+i\Delta_2)/[(1+i\Delta_1)(\alpha+i\Delta_2)+I_1]$ and $B_0 = i\kappa^*A_0^2/(\alpha+i\Delta_2)$ that satisfy
\begin{equation}
	I_1\left[(\Delta_1\Delta_2-\alpha-I_1)^2+(\alpha\Delta_1+\Delta_2)^2\right] = P(\alpha^2+\Delta_2^2),
	\label{eq:cw}
\end{equation}
where we have introduced the notation $I_1 = |\kappa|^2|A_0|^2$, $I_2 = |\kappa|^2|B_0|^2$ and $P = |\kappa|^2|S|^2$. The solutions are stable attractors for an initially empty cavity system in the presence of a low power driving field. The homogeneous solution can display bistability, given that the two conditions $\Delta_1\Delta_2 > \alpha$ and $|\Delta_2|(|\Delta_1|-\sqrt{3})/(\sqrt{3}|\Delta_1|+1) > \alpha$ are satisfied \cite{Etrich_1997a}, which also requires the detunings to have the same sign.

The generation of subharmonic sidebands that can seed the formation of a frequency comb occurs when the homogeneous solution becomes modulationally unstable \cite{Trillo_1996}. A linear stability analysis assuming a perturbation of the form $A = A_0 + a_1 e^{\lambda t+i\Omega\tau} + a_{-1} e^{\lambda^* t-i\Omega\tau}$ (and analogous for $B$) gives a characteristic equation for the potentially growing eigenvalues
\begin{equation}
	\left[(\lambda+1)^2+f_1\right]\left[(\lambda+\overline{\alpha})^2+f_2\right] = p,
	\label{eq:char}
\end{equation}
where $f_1 = \overline{\Delta}_1^2 + 2I_1-I_2$, $f_2 = \overline{\Delta}_2^2 + 2I_1$ and $p = 2I_1\left[(1-\overline{\alpha})^2+(\overline{\Delta}_1+\overline{\Delta}_2)^2-I_2\right]$, and we have introduced $\overline{\alpha} = \alpha+id\Omega$, $\overline{\Delta}_1 = \Delta_1-\eta_1\Omega^2$ and $\overline{\Delta}_2 = \Delta_2-\eta_2\Omega^2$. 

\begin{figure}[htb]
\centerline{\includegraphics[width=\linewidth]{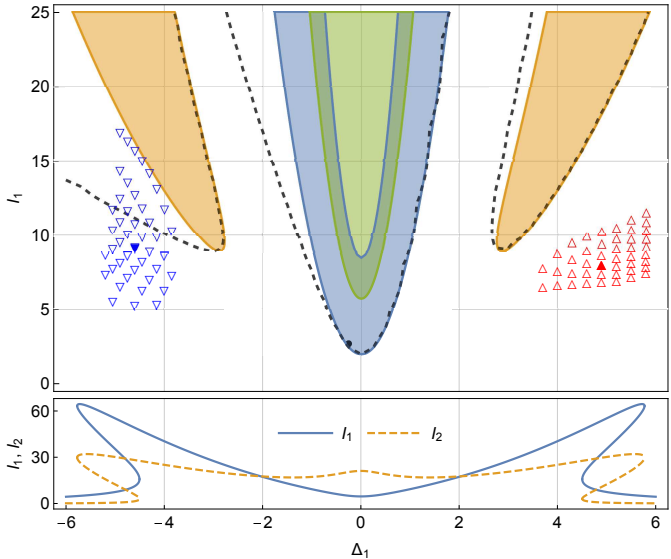}}
\caption{Main panel: Stability phase diagram for $\alpha = 1$ and $\Delta_2 = 2\Delta_1$. The homogeneous solutions are unstable to cw perturbations within the shaded regions and modulationally unstable within the dashed contour. Bottom panel: Variation in FF/SH intracavity power for the homogeneous solution with $S = 12$.}
\label{fig:loci}
\end{figure}

In order to guide the search for soliton solutions in the parameter space, we perform an analysis of the SHG systems critical points. The conditions for which a growing instability will develop can be determined from Eq.~(\ref{eq:char}) by application of the Routh-Hurwitz stability criteria. The homogeneous solution will, in the absence of walk-off (i.e. for $d = 0$), become unstable if either of the following inequalites are satisfied: i) the condition $f_1+\alpha f_2 + (1+\alpha)[\alpha+(1+\alpha)^2] < 0$ (green region in Fig.~\ref{fig:loci}), ii) $\alpha(f_1-f_2) + [p+2\alpha(f_1+f_2)](1+\alpha)^2 + \alpha(1+\alpha)^4 < 0$ (blue region) and iii) $(1+f_1)(\alpha^2+f_2) - p < 0$ (orange regions). For the case under consideration when the loss ratio is unity, we further have that $\overline{\alpha} = 1$, so that the characteristic equation is biquadratic and can be solved explicitly, to give the eigenvalues
\begin{equation}
	\lambda = -1 \pm \sqrt{-\frac{1}{2}(f_1+f_2) \pm \sqrt{p+\frac{1}{4}(f_1-f_2)^2}}.
\end{equation}
Figure \ref{fig:loci} shows the stability phase diagram in the parameter space $(\Delta_1,I_1)$, where the different instability regions for perturbations with $\Omega = 0$ are colored. Each point on the diagram corresponds to a comb state that can be realized by the stationary homogeneous solution for some particular combination of pump power and detuning. The orange shaded regions mark domains that correspond to the unstable middle branch of the bistable homogeneous solution, while the system has complex conjugated eigenvalues and may display self-pulsing \cite{McNeil_1978} due to a Hopf bifurcation within the blue shaded region. The blue and green regions are seen to partially overlap, with the eigenvalues ceasing to be oscillatory at the upper boundary of the blue domain and becoming purely real within the non-overlapping green region. The boundaries of the domains that exhibit modulational instability (MI) to periodic perturbations are additionally marked in the figure with a dashed contour. The MI domains are asymmetric, owing to the different phase matching contributions that come from the quantities $\overline{\Delta}_{1,2}$ for positive and negative detuning. The power for the FF is seen to have a minimum threshold for the onset of an instability that occurs for $\Delta_1 = \Delta_2 = 0$ and is given by $I_1^{th} = \alpha(1+\alpha)$.

Having established the stability properties of the homogeneous solution, we proceed to numerically search for localized structures. We limit our search for bright CSs by considering parameter regions near which the homogeneous solution displays a bistability such that the lower branch is stable, whereas the upper branch is modulationally unstable to perturbations with a finite periodicity. In general we may expect to find CSs in parameter regions with coexistence of two different stable stationary states where: i) both states may be homogeneous, or ii) one is homogeneous while the other is a periodic pattern. The CSs solutions can then form due to the locking of fronts connecting both states. 

To find the stationary solutions we employ a numerical Newton-Raphson method that solves Eqs.~(\ref{eq:S1}-\ref{eq:S2}) with the slow-time ($t$) derivative set to zero. The localized solutions are excited using a Gaussian writing pulse of variable amplitude and width that is added to the pump field during the first few iterations of the solver and later removed. We also verify the stability of the solutions under propagation using a split-step Fourier method that integrates Eqs.~(\ref{eq:S1}-\ref{eq:S2}) with a 4th-order Runge-Kutta scheme for evaluation of the nonlinear step. All simulations are made assuming a fixed frequency spacing with $N = 2048$ modes and a normalized FSR of $1/250$.

\begin{figure}[htb]
\centerline{\includegraphics[width=\linewidth,trim={1.8cm .1cm 1cm .9cm},clip]{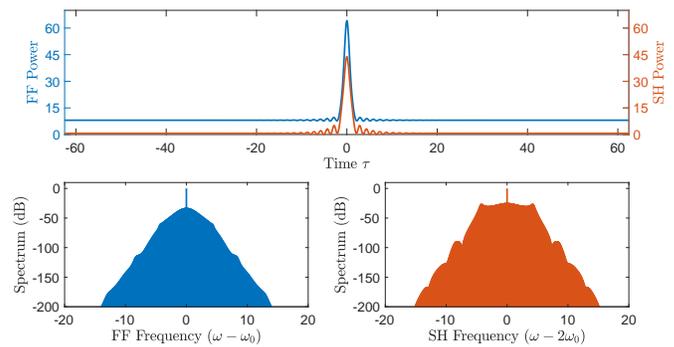}}
\caption{Normalized intracavity power (top) and spectra (bottom) of a bright cavity soliton for driving strength $S = 12$ and detuning $\Delta_1 = 4.9$ (marked by a filled red upwards-facing triangle in Fig.~\ref{fig:loci}).}
\label{fig:CS1}
\end{figure}

We first search for bright solitons beyond the bistability threshold for positive detuning. An example of such a solution is shown in Fig.~\ref{fig:CS1}. Both the FF and SH amplitudes are seen to display a localized central peak with damped oscillations on either side. The respective spectra also exhibit a fine structure with a modulated envelope shape. We emphasize that the signs for the GVD of the two fields differ, and we find that the solitons corresponds to a localized portion of a pattern embedded in the homogeneous background. The location of a number of these solutions are marked by red upwards-facing triangles in the phase diagram shown in Fig.~\ref{fig:loci}, where the y-axis indicates the power of the corresponding FF background. The CSs are found to connect to a family of stationary Turing patterns for decreasing detuning value, and a bifurcation analysis shows that the pattern is subcritical with the CSs undergoing a homoclinic snaking bifurcation structure \cite{PPR_2018}. We have verified that the quadratic solitons indeed behave and have the properties that are usually associated with cavity solitons. The solitons are robust to perturbations and have a unique amplitude and width for a given set of pump parameters. Multiple solitons are also non-interacting for sufficiently large separation, and are individually addressable, so that they can be written and erased, by the addition of a writing pulse that is either in or out of phase with the pump field. We have also estimated the power needed for the soliton formation. The driving strength is related to the pump field amplitude as $S = \sqrt{\theta_1}A_{in}\hat{\kappa} L/\alpha_1^2$ (see Ref.~\cite{Leo_2016b}). Assuming critical coupling $\theta_1 = \alpha_1$, a cavity finesse $\mathcal{F} = \pi/\alpha_1 = 160$, a nonlinear coefficient $\hat{\kappa} = 11.14~\textrm{W}^{-1/2}\textrm{m}^{-1}$ and a crystal length $L = 15~\textrm{mm}$ (c.f.~Refs.~\cite{Ricciardi_2015,Leo_2016a}), we find that a driving strength of $S = 12$ corresponds to a mere $39~\textrm{mW}$ of pump power.

An important property of SHG systems is that they are generally insensitive to the sign of the dispersion, in the sense that a given system will display similar dynamics for normal dispersion and positive detuning as it does for anomalous dispersion and negative detuning \cite{Hansson_2017}. Indeed, the eigenvalues of Eq.~(\ref{eq:char}) are invariant to a simultaneous sign reversal of detunings and GVDs. In fact, the use of quadratic nonlinearities may enable us to realize solutions that are characteristic for both anomalous and normal dispersion in the very same system. In particular, given that we have found bright CSs for positive detuning, we may expect to find dark (or gray) CSs for negative detuning.

\begin{figure}[htb]
\centerline{\includegraphics[width=\linewidth,trim={1.8cm .1cm 1cm .9cm},clip]{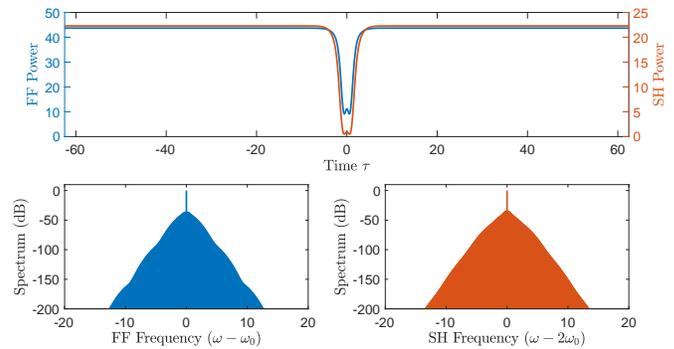}}
\caption{Example of a stable localized dark soliton solution for driving strength $S = 10$ and detuning $\Delta_1 = -4.6$ (marked by a filled blue downwards-facing triangle in Fig.~\ref{fig:loci}).}
\label{fig:CS2}
\end{figure}

A typical example of a dark localized structure is shown in Fig.~\ref{fig:CS2}. The predicted locations of a number of dark solutions have also been marked by blue downwards-facing triangles in Fig.~\ref{fig:loci}, where the y-axis now indicates the minimum power attained in the dip of the FF temporal profile. The dark CSs correspond to holes in the modulationally stable upper branch homogeneous solution, where the intracavity power is locally reduced. The dark soliton is seen to exhibit a small bump in the center where two interlocking fronts connect the bistable homogeneous solution with a periodic pattern on the lower branch. The soliton in Fig.~\ref{fig:CS2} is the narrowest structure that allows for the interlocking of the two fronts, but we note that it is also possible to find dark soliton solutions of different widths that include multiple periods of the embedded pattern.

Although these results have been obtained for a medium with mixed dispersion, we remark that it is also possible to find stable bright soliton solutions when the two GVD coefficients have equal signs. This corresponds to the same case for which solitons have previously been studied in 2D diffractive SHG systems \cite{Etrich_1997a}. 

It is clear that quadratic CSs display some similarities with Kerr frequency combs. 
To better understand this behavior, it can be helpful to consider a simplified model using the reduced single envelope equation for the FF that was previously derived in Ref.~\cite{Leo_2016b}. Here, one finds that the cascading of the three-wave mixing process can be described by an effective cubic nonlinearity with a non-instantaneous nonlinear response function. The reduced equation can, in the instantaneous (local) limit, be shown to be formally equivalent to a Lugiato-Lefever model with a nonlinear coefficient proportional to $-\alpha+i\Delta_2$ that includes nonlinear loss. We can consequently see that it is the sign of the SH detuning ($\Delta_2$) in combination with the sign of the FF GVD ($\eta_1$) that determines whether the comb dynamics correspond to an effective normal or anomalous dispersion. Moreover, this analysis shows that the localized structures of Eqs.~(\ref{eq:S1}-\ref{eq:S2}) may be connected with Kerr solitons of the Lugiato-Lefever equation, and that these are able to persist in the presence of weak nonlinear loss and a non-instantaneous nonlinearity.

\begin{figure}[htb]
\centerline{\includegraphics[width=\linewidth,trim={1.7cm 1cm 2.6cm 1.5cm},clip]{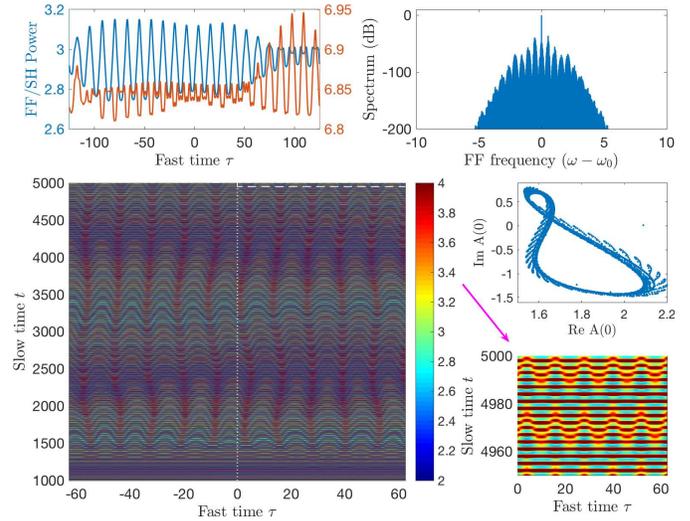}}
\caption{Turbulent evolution of intracavity power for $S = 6$ and $\Delta_1 = -0.3$ ($I_1 = 3.07$). Bottom left: Contour plot showing long-term evolution of FF power. Top left: Intracavity power of FF/SH (blue/red) fields at $t = 5000$. Top right: Normalized FF spectrum at $t = 5000$. Middle right: Characteristic figure 8 field portrait along $\tau = 0$ (dotted line). Bottom right: Zoom-in showing evolution on a shorter time scale inside the region marked by a dashed white line.}
\label{fig:turb}
\end{figure}

Finally, we consider the possibility of comb generation due to the self-pulsing (Hopf) instability \cite{McNeil_1978} that occurs for low powers and small detuning values within the blue shaded region of Fig.~\ref{fig:loci}. Here we find that the MI growth rate for perturbations with a finite periodicity will generally exceed (lose to) the growth rate of homogeneous perturbations when the detuning is negative (positive). The nonlinear evolution of perturbations with $\Omega = 0$ manifests itself as a periodic oscillation, or a self-pulsing, of the homogeneous background that occurs on the slow time-scale. Meanwhile, the simultaneous growth of perturbations with $\Omega \neq 0$ lead to the appearance of temporal oscillations that occur on the fast time-scale. Although the unstable eigenvalues are complex, it may be possible to generate stationary pattern structures within this region \cite{Trillo_1996}. However, for the parameter values corresponding to the mixed dispersion configuration, we find that the comb evolution generally displays a turbulent and chaotic behaviour without a steady-state. This can be seen in Fig.~\ref{fig:turb}, which shows the long-term evolution of the intracavity power at a point (marked by a black dot in Fig.~\ref{fig:loci}) inside the instability boundary. These combs have a very low threshold power, a relatively broad bandwidth, and a single FSR comb line spacing. They exhibit some systematic features, but are characterized by a lower degree of coherence than combs of the cavity soliton type. 

In conclusion, we have investigated the generation of quadratic frequency combs in a doubly-resonant SHG cavity system for which the temporal walk-off can be made to vanish. We have reported conditions for which comb generation can occur, and identified some dynamic regimes in which both bright and dark localized CS solutions may be observed in the same system. These are found to have analogous properties to Kerr CSs, which suggests that quadratic frequency combs may indeed be a viable alternative to Kerr combs, that can offer unique benefits for a variety of applications.

\textbf{Funding.} The research leading to these results has received funding from the European Union's Horizon 2020 research and innovation programme under the Marie Sk\l{}odowska-Curie grant agreement No GA-2015-713694. Additionally, T.H. acknowledges funding from the Swedish Research Council (Grant No. 2017-05309); PPR acknowledge support from the internal Funds from KU Leuven; M.B. and S.W. acknowledges funding by the Italian Ministry of University and Research (MIUR) (Project PRIN 2015KEZNYM - NEMO); F.L. acknowledges funding from the European Research Council (ERC) under the European Union's Horizon 2020 research and innovation programme (Grant agreement No. 757800). 

 


\end{document}